\documentclass[reprint,superscriptaddress,amsmath,amsfonts,amssymb,aps,prl,showkeys]{revtex4-1}
\usepackage{bm}
\usepackage{color}
\usepackage{graphicx}
\usepackage[normalem]{ulem}

\usepackage{hyperref}
\hypersetup{
  breaklinks = {true},
  citecolor = {blue},
  colorlinks = {true},
  linkcolor = {red},
  urlcolor = {blue},
}

\graphicspath{{../Figs/}}

\begin{document}

\title{Valley-Polarized Quantum Anomalous Hall Phase in Bilayer Graphene \\ with Layer-Dependent Proximity Effects}

\author{Marc Vila}
\affiliation{Catalan Institute of Nanoscience and Nanotechnology (ICN2), CSIC and BIST, Campus UAB, Bellaterra, 08193 Barcelona, Spain}

\author{Jose H. Garcia}
\affiliation{Catalan Institute of Nanoscience and Nanotechnology (ICN2), CSIC and BIST, Campus UAB, Bellaterra, 08193 Barcelona, Spain}

\author{Stephan Roche}
\affiliation{Catalan Institute of Nanoscience and Nanotechnology (ICN2), CSIC and BIST, Campus UAB, Bellaterra, 08193 Barcelona, Spain}
\affiliation{ICREA--Instituci\'o Catalana de Recerca i Estudis Avan\c{c}ats, 08010 Barcelona, Spain}

\date{\today}

\begin{abstract}
Realizations of some topological phases in two-dimensional systems rely on the challenge of jointly incorporating spin-orbit and magnetic exchange interactions. Here, we predict the formation and control of a fully valley-polarized quantum anomalous Hall effect in bilayer graphene, by separately imprinting spin-orbit and magnetic proximity effects in different layers. This results in varying spin splittings for the conduction and valence bands, which gives rise to a topological gap at a single Dirac cone. The topological phase can be controlled by a gate voltage and switched between valleys by reversing the sign of the exchange interaction. By performing quantum transport calculations in disordered systems, the chirality and resilience of the valley-polarized edge state are demonstrated. Our findings provide a promising route to engineer a topological phase that could enable low-power electronic devices and valleytronic applications as well as putting forward layer-dependent
proximity effects in bilayer graphene as a way to create versatile topological states of matter.
\end{abstract}

\maketitle

%

Topological phases of matter hold great potential in a myriad of fields such as low-power electronics, spintronics, sensing, metrology or quantum information processing \cite{Nayak2008, Hasan2010, Qi2011, Ren2016, Liu2016}. The quantum anomalous Hall effect (QAHE) is one of such phases \cite{Haldane1988, Ohgushi2000, Onoda2003}, displaying dissipationless chiral edge states, quantized Hall conductivity and finite Chern number at zero external magnetic field \cite{Thouless1982, Hatsugai1993}. Since topological currents are extremely robust to disorder effects \cite{Hasan2010}, they constitute a very attractive platform for ultralow power electronics. Many materials can host the QAHE, such as semiconductor quantum wells \cite{Liu2008, Wang2014}, graphene \cite{Qiao2010, Tse2011, Ding2011, Qiao2012, Offidani2018, Hogl2020} and other two-dimensional systems \cite{Ezawa2012, Zou2020}, or more recently magnetic topological insulators \cite{Yu2010, Chang2013, Bestwick2015, Otrokov2019, Tokura2019, Deng2020} and twisted bilayer graphene \cite{Serlin2020, Chen2020, Polshyn2020, Tschirhart2021}.
Among the vast playground of quantum materials \cite{Giustino2021}, graphene-based compounds are very promising due to their chemical stability, multiple degrees of freedom (lattice, spin, valley), ease of device fabrication and scalability as well as the possibility to combine them with other materials to form van der Waals heterostructures \cite{Geim2013, Sierra2021}. Unfortunately, the main ingredient of the QAHE is the coexistence of spin-orbit coupling (SOC) and magnetism \cite{Ren2016, Liu2016}, and these effects are generally very small or absent in graphene-based devices. Accordingly, engineering these interactions via proximity effects \cite{Zutic2019, Huang2020} has been a major focus during the last decade \cite{Garcia2018, Roche2015, Yang2013, Hallal2016, Wang2015, Wei2016, Singh2017, Leutenantsmeyer2017, Karpiak2019, Gmitra2015, Gmitra2016, Jin2013, Song2018, Benitez2017, Ghiasi2017, Khokhriakov2018, Karpiak2019}. Nevertheless, while adding either SOC or exchange has been reported experimentally (with characteristic spin splittings and spin textures), imprinting both effects simultaneously and achieving a measurable QAHE still remains a tantalizing challenge. This would require encapsulating graphene between a ferromagnetic insulator (FMI) and a strong-SOC material, as proposed recently in Ref. \cite{Zollner2020}, or to use a single material that can induce both interactions simultaneously \cite{Qiao2014, Zhang2015}.

\begin{figure}[tb]
\includegraphics[width=0.47\textwidth]{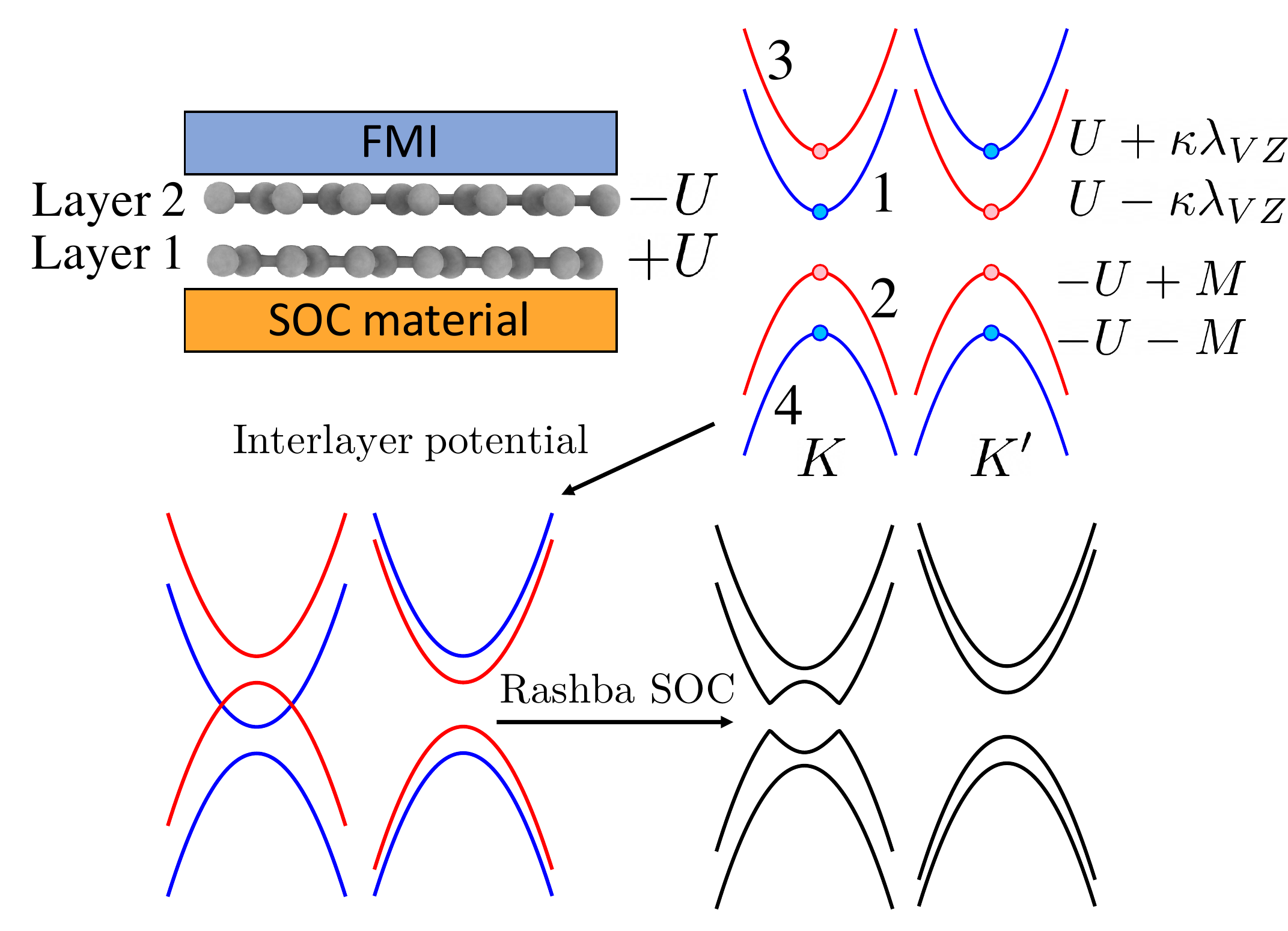}
\caption{Top left: Van der Waals heterostructure comprised of a SOC material, bilayer graphene and a ferromagnetic insulator. Top right and bottom: Mechanism  of the valley-polarized quantum anomalous Hall effect. Red and blue colors depict up and down spins, respectively.}
\label{fig_F1}
\end{figure}

In addition to SOC and magnetism, the layer degree of freedom of bilayer graphene (BG) and the possibility to generate a sizable band gap by applying an external perpendicular electric field offers an additional knob to engineer topological phases \cite{Tse2011, Qiao2013, Alsharari2018, Island2019, Zaletel2019, Tiwari2021}. Unfortunately, the proposed models for realizing the QAHE in bilayer graphene have been limited to some extension of the monolayer graphene case where both layers require SOC and exchange \cite{Tse2011, Qiao2013}. Recently, proximity effects on one of the layers in BG was shown to mainly affect the conduction (or valence) band owing to the layer localization of these low energy states \cite{Gmitra2017, Khoo2017, Zollner2018, Alsharari2018, Cardoso2018, Tiwari2021PRL, Ingla2021, McCann2006, McCann2013}. Accordingly, such effect can be harnessed to add SOC and magnetism separately on each layer by sandwiching BG between SOC and magnetic materials \cite{Zollner2020}, as sketched in Fig. \ref{fig_F1}. In such device configuration, the wave function localization can be reversed by interchanging the interlayer potential with a perpendicular electric field, hence enabling a possible swapping of the induced spin splittings between the conduction and valence bands and therefore modulate the proximity effects.  

In this Letter, we take advantage of this layer-dependent proximity effect to predict the formation of a quantum anomalous Hall phase in encapsulated bilayer graphene. By introducing SOC and exchange interactions separately in different layers, the valence and conduction bands hybridize in such a way that a topological gap opens but only at a single Dirac cone, hence resulting in a valley-polarized quantum anomalous Hall effect (VP-QAHE) with Chern number $\mathcal{C} = \pm 1$, previously only theorized in silicene \cite{Pan2014, Pan2015} and in artificial honeycomb lattices \cite{Zhou2017}. This proposal of QAHE is likely more experimentally feasible than previous mechanisms \cite{Tse2011, Qiao2013} since only one interaction per layer is needed, suggesting that heterostructures with common materials such as two-dimensional magnets/BG/transition metal dichalcogenides could be used. We show that the valley-polarization and chirality of the edge states is reversed by changing the sign of the exchange interaction, and that an applied electric field can switch the topological phase.
Transport calculations in nonlocal geometries further reveal that the valley-polarized edge states are robust to both bulk and edge disorders. Finally, we argue that the VP-QAHE should be observable with currently employed materials and discuss possible applications.

Our starting point is the tight-binding of bilayer graphene encapsulated between a ferromagnetic insulator and a strong-SOC material (Fig. \ref{fig_F1}). The Hamiltonian is that of Bernal-stacked BG \cite{McCann2013} with SOC and exchange interaction, namely, $\mathcal{H} = \mathcal{H}_0 + \mathcal{H}_\text{SOC} + \mathcal{H}_\text{ex}$. The first term is the orbital part $\mathcal{H}_0 = -t \sum_{\langle i,j \rangle, s} c^\dagger_{is} c_{js} + t_\perp \sum_{\langle i\in A_1, j\in B_2 \rangle, s} c^\dagger_{is} c_{js} + U \sum_{i,s} \eta_i c^\dagger_{is}c_{is}$. Each term is detailed elsewhere \cite{H0, Suppmat} and here we just emphasize that $U$ describes an interlayer potential that opens a gap of magnitude $2U$ at the Dirac point (for $U \ll t_\perp$) \cite{McCann2013}.
For $\mathcal{H}_\text{SOC}$ we assume the typical terms induced on graphene by a transition metal dichalcogenide \cite{Gmitra2015, Gmitra2016, Kochan2017} or topological insulators \cite{Song2018, Zollner2020TI}, which includes Rashba and sublattice-dependent intrinsic SOC:
\begin{align}
\mathcal{H}_\text{SOC} &= \frac{2i\lambda_R}{3} \sum_{\langle i,j \rangle, s,s^\prime}c^\dagger_{is} [(\bm{s} \times \bm{d}_{ij}) \cdot \hat{\bm{z}}]_{ss^\prime} c_{js^\prime}  \nonumber \\
&+ \frac{i}{3\sqrt{3}} \sum_{\langle\langle i,j \rangle\rangle, s} \lambda_I^i \nu_{ij} c^\dagger_{is} s_z c_{js}.
\end{align}
The first line describes the spin-mixing Rashba SOC with strength $\lambda_R$, while $\bm{s} = (s_x, s_y, s_z)$ is a vector of spin Pauli matrices and $\bm{d}_{ij}$ a unit vector from site $j$ to $i$. The other term is the spin-conserving intrinsic SOC with value $\lambda_I^A$ ($\lambda_I^B$) if the hopping connects atoms of the sublattice A (B); and $\nu_{ij} = \pm 1$ if the hopping is counterclockwise (clockwise). We can rewrite the intrinsic SOC as the Kane-Mele SOC \cite{Kane2005}, $\lambda_{KM} = (\lambda_I^A + \lambda_I^B)/2$, and valley-Zeeman SOC \cite{Gmitra2015, Gmitra2016, Kochan2017}, $\lambda_{_{VZ}} = (\lambda_I^A - \lambda_I^B)/2$. $\lambda_{KM}$ opens a topological gap at K/K$^\prime$ valleys whereas $\lambda_{_{VZ}}$ produces an exchange field with opposite signs at different valleys. Here, we set $\lambda_{KM} = 0$ as this term is always much smaller than any other parameter \cite{Garcia2018}. Finally, the magnetic proximity effect is modeled by the term
\begin{equation}
\mathcal{H}_\text{ex} = M \sum_{i,s,s^\prime} c^\dagger_{is} [\bm{m}\cdot \bm{s}]_{ss^\prime}  c_{is^\prime},  
\end{equation}
where $\bm{m} = \hat{\bm{z}}$ is the magnetization direction of the FMI and $M$ the exchange coupling.

The mechanism proposed in this work is illustrated in Fig. \ref{fig_F1}. Thanks to the layer-dependent proximity effects, the valence band is spin-splitted due to the exchange coupling $Ms_z$, while $\kappa \lambda_{_{VZ}}s_z$ dictates the conduction band splitting (with $\kappa = \pm 1$ for K and K$^\prime$, respectively). For illustration purpose, the system initially presents a finite gap due to a substrate-induced interlayer potential $U$ \cite{Gmitra2017}. Then, tuning the $U$ value with a perpendicular electric field allows to reduce such band gap and makes the conduction and valence band overlap. Since the spins of the conduction band minimum and valence band maximum at K$^\prime$ are identical, the bands interchange but the gap remains nonzero \cite{Liu2008}. However, the spins are opposite at the K valley, resulting in a band crossing. Upon the addition of the Rashba SOC, $s_z$ is not a good quantum number anymore due to spins mixing, and the band hybridization yields a gap opening \cite{Yu2010, Qiao2010}.
We argue below that this gap is topological and displays the VP-QAHE.

\begin{figure}[tb]
\includegraphics[width=0.47\textwidth]{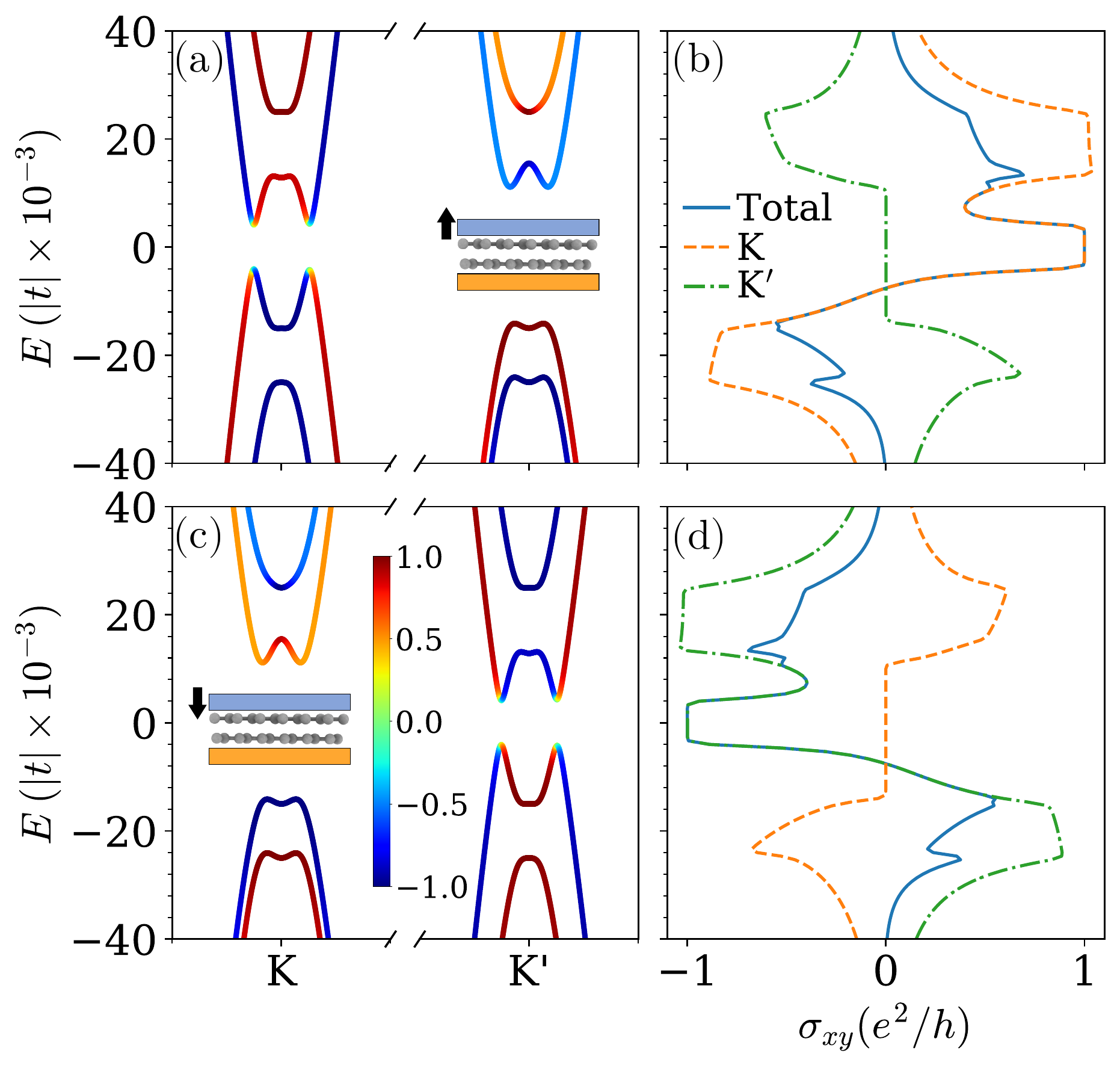}
\caption{(a) Band structure of $\mathcal{H}$ near the Dirac cones. The band color depicts the $z$ component of the spin texture. (b) Total and valley-resolved Hall conductivities of the system in (a), at zero temperature. (c) and (d) are the same as (a) and (b) but for reversed exchange interaction. The tight-binding parameters in units of $t$ are: $t_\perp = 0.13$, $\lambda_{_{VZ}} = 0.015$, $\lambda_R = 0.015$, $M = \pm 0.025$, $U = 0$.} 
\label{fig_F2}
\end{figure}

\begin{figure}[tb]
\includegraphics[width=0.47\textwidth]{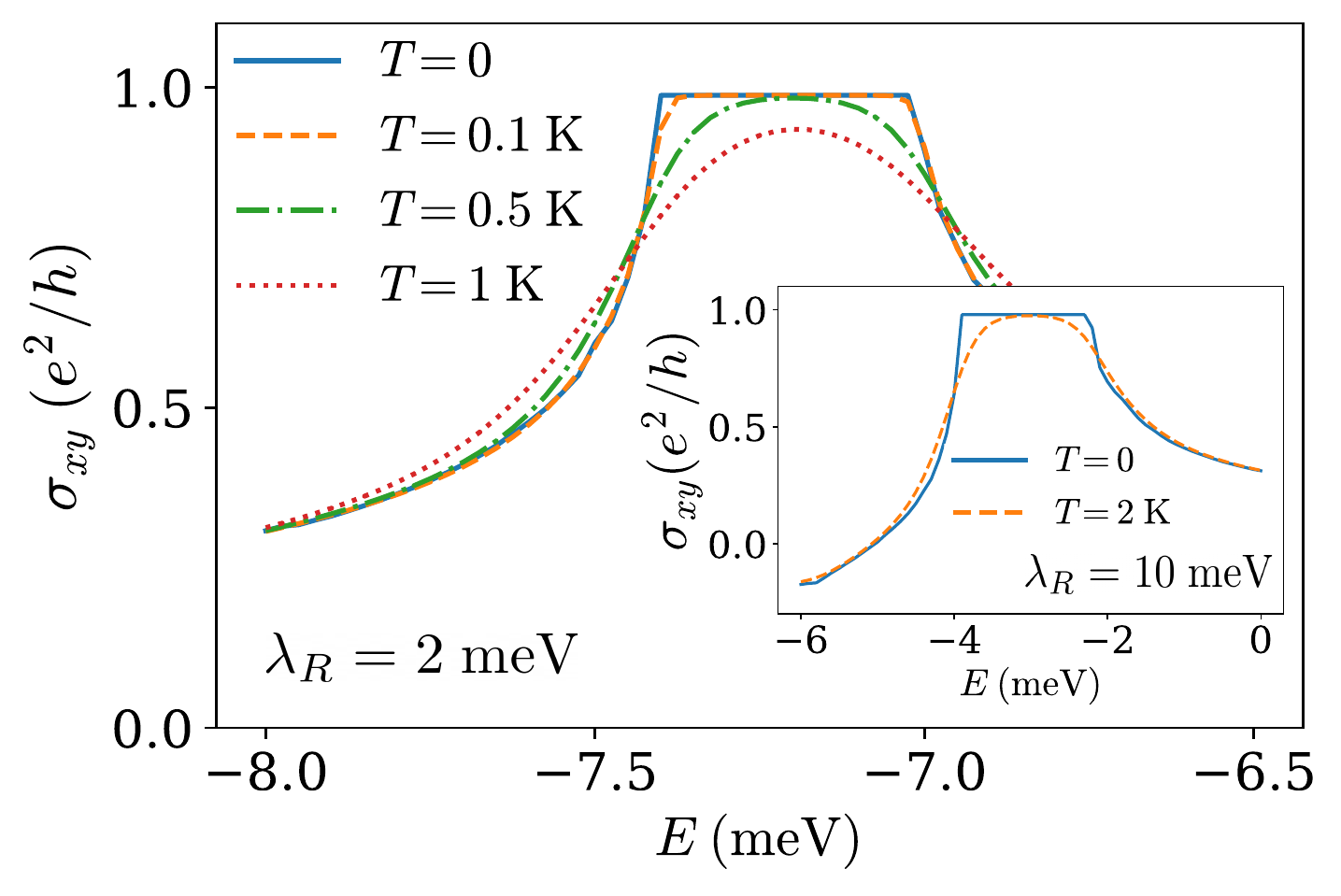}
\caption{Hall conductivity for realistic model parameters at different temperatures $T$. The tight-binding parameters are: $t=2.7$ eV, $t_\perp = 0.339$ eV, $\lambda_{_{VZ}} = \lambda_R = 2$ meV, $M = 30$ meV, $U = -9$ meV. For the inset, $\lambda_R =10$ meV and $U = -4$ meV.} 
\label{fig_F5}
\end{figure}

Fig. \ref{fig_F2}(a) shows the band structure and spin texture near the Dirac cones once the bands have hybridized, noting the usage of large parameters to emphasize the qualitative features of the model, while more realistic parameters are later discussed in relation to experiments (see Fig. \ref{fig_F5}). To confirm the presence of the QAHE, we compute the Chern number $\mathcal{C}=\frac{1}{2\pi}\int \bm{\Omega}(\bm{k}) \cdot \text{d}^2\bm{k}$, whose value determines the number of topological edge states and relates to the intrinsic Hall conductivity as $\sigma_{xy}=\mathcal{C} \frac{e^2}{h}$ \cite{Thouless1982}. The quantity $\bm{\Omega}(\bm{k})$ is the Berry curvature and for a two-dimensional system equals to \cite{Xiao2010Berry, Yao2004}
\begin{equation}
\small
\bm{\Omega}(\bm{k}) = -2\hbar^2 \sum_n f_n    \sum_{m\neq n} \text{Im} \left\{ \frac{\langle n,\bm{k} | v_x | m,\bm{k}\rangle \langle m,\bm{k} | v_y | n,\bm{k} \rangle }{(E_n (\bm{k}) - E_m (\bm{k}))^2} \right\}.
\end{equation}
Here, $n$ and $m$ are band indices, $|nk\rangle$ is a Bloch state with energy $E_n (\bm{k})$, $f_n$ is the Fermi-Dirac distribution and $v_{i=x,y} = \partial \mathcal{H}/\hbar\partial k_i$ is the velocity operator.

Our system displays nonzero Berry curvature only close to K and K$^\prime$-points. Consequently, we can perform the integration of the Berry curvature near each valley separately and obtain the valley-resolved Hall conductivity together with its total value; we plot these quantities in Fig. \ref{fig_F2}(b). In the energies corresponding to the band gap, the total Hall conductivity is quantized at exactly $e^2/h$, proving the existence of the QAHE with $\mathcal{C}=1$. Importantly, the valley-resolved conductivities reveal that all the contribution comes only from the K valley, indicating that the gap at K (K$^\prime$) is topological (trivial) and that the QAHE is valley-polarized. This fact can also be inferred from the spin texture because the $z$ component of the spin changes sign abruptly at the anticrossing point of the topological gap.

Furthermore, changing the sign of either $\bm{m}$ or $\lambda_{_{VZ}}$ will swap the topology of each valley. Experimentally, $\lambda_{_{VZ}}$ is fixed by the interaction of graphene with the substrate, but $\bm{m}$ can be reversed easily by a magnetic field or even by electric means in multiferroic systems \cite{Matsukura2015} or two-dimensional magnets \cite{Huang2018, Jiang2018, Wang2018, Johansen2019}. Figs. \ref{fig_F2}(c) and \ref{fig_F2}(d) show the bands and Hall conductivity for $\bm{m} \rightarrow -\bm{m}$, respectively. One clearly sees that the band structure is exactly as opposite as that of panel (a), and $\sigma_{xy}$ indeed confirms the opening of the topological gap at K$^\prime$. Interestingly, not only does the valley polarization of the Hall conductivity change, but also its sign, indicating that the chirality of the edge states is reversed. In contrast, if $\bm{m}$ is kept fixed but it is the sign of $\lambda_{_{VZ}}$ that is inverted (not shown), the valley polarization also flips while $\sigma_{xy}$ remains unchanged. 

\begin{figure}[tb]
\includegraphics[width=0.44\textwidth]{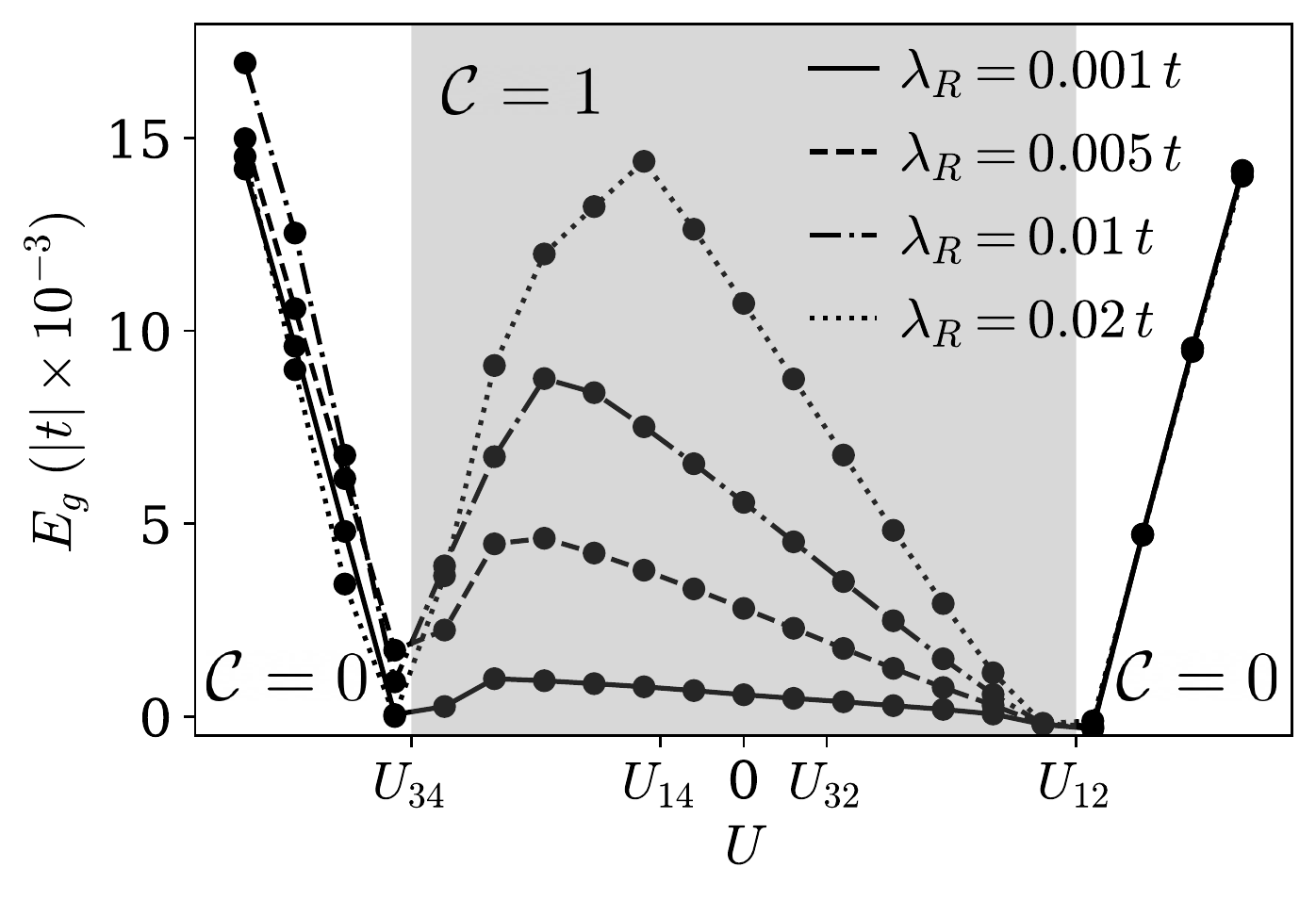}
\caption{Band gap as a function of interlayer potential for different Rashba strengths. Gray shading shows the region with nonzero Chern number. The values of $U_{ij}$ are detailed in the main text. Parameters other than $U$ and $\lambda_R$ are kept the same as in Fig. \ref{fig_F2}.} 
\label{fig_F3}
\end{figure}

It is not obvious whether the system remains topological if $U$ varies and more bands hybridize. Therefore, we study how the topological phase and the band gap value, $E_g$, depend on the model parameters. For that, it is useful to label the four low-energy bands, as shown by the numbers 1-4 in Fig. \ref{fig_F1}. Then, assuming the spin splitting induced by exchange is larger than that of SOC, the critical values of $U=U_{ij}$ that make bands $i$ and $j$ overlap at K/K$^\prime$ can be evaluated analytically: $U_{12} = (|M| + 	|\lambda_{_{VZ}}|)/2$, $U_{32} = (|M| - |\lambda_{_{VZ}}|)/2$, $U_{14} = - U_{32}$ and $U_{34} = - U_{12}$. These values are actually approximate because $\lambda_R$ will modify the band dispersion, but they are nevertheless in very good agreement with full numerical calculations. Fig. \ref{fig_F3} displays $E_g$ and $\mathcal{C}$ as a function of $U$ for different Rashba strengths. One clearly observes that the topological phase transitions occur only at $U_{12}$ and $U_{34}$. Consequently, the VP-QAHE persists in the range $U = U_{12} - U_{34} = |M| + |\lambda_{_{VZ}}|$. This entails that increasing the proximity interactions facilitates the range of electric fields needed to access the topological phase. Moreover, the maximum value of the band gap is neither centered at $U=0$ nor is the same for distinct $\lambda_R$. From the different plots of Fig. \ref{fig_F3}, it is seen that $E_g$ is proportional to $\lambda_R$ in the topological phase. The scaling is approximately linear in the range of values studied, while larger Rashba SOC leads to sublinear scaling and even gap closing (not shown here) since $\lambda_R$ more strongly distorts the bands, hence invalidating the mechanism described in Fig. \ref{fig_F1}.

\begin{figure}[tb]
\includegraphics[width=0.47\textwidth]{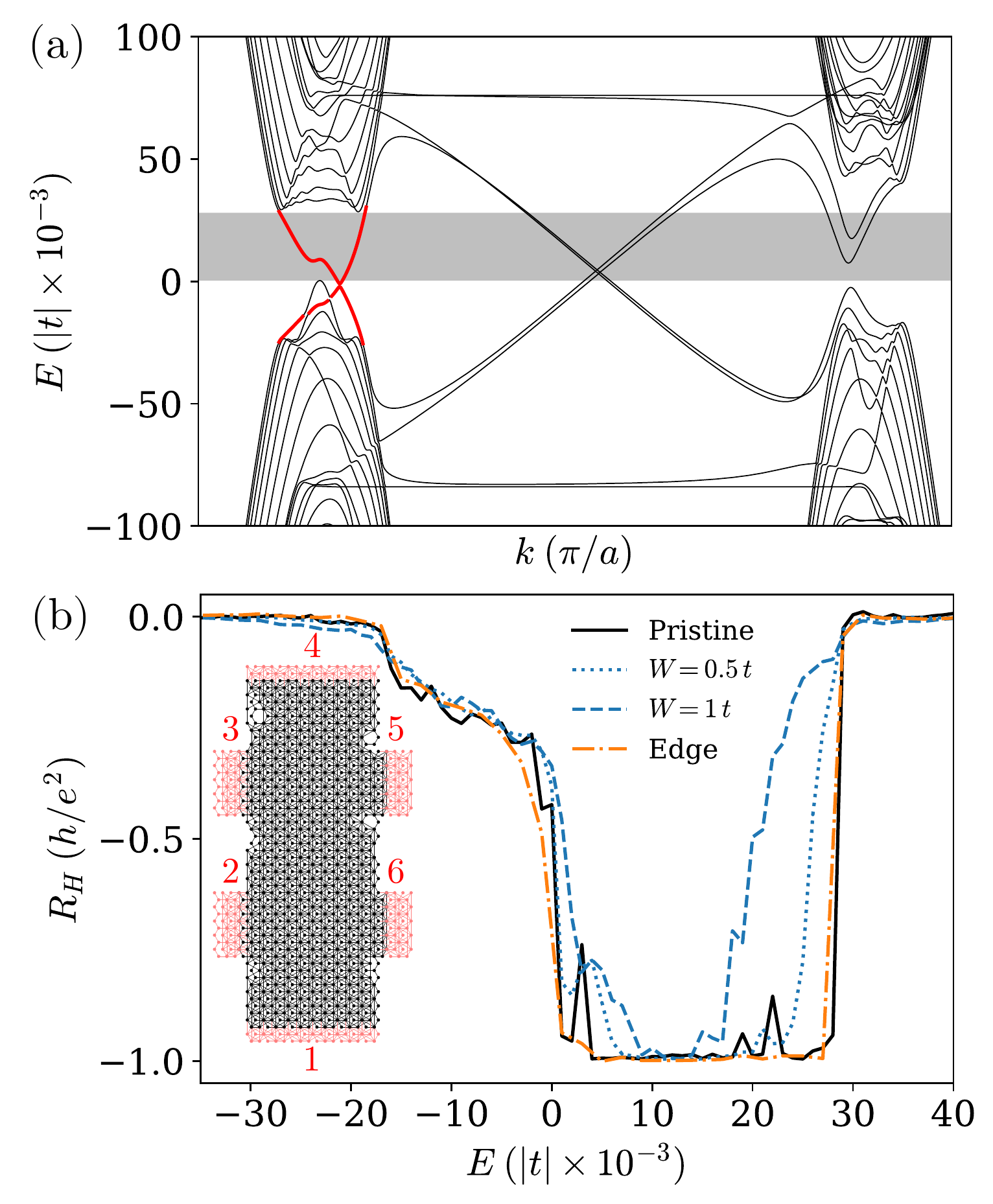}
\caption{(a) Band structure of a zigzag nanoribbon of width 25 nm. The gray shading and red bands denote the energies in (b) where the Hall resistance is quantized and the topological edge states, respectively. (b) Hall resistance of a 6-terminal device for different types of disorder, averaged over 50 random disorder configurations. Inset: schematics of the 6-terminal device with disordered edges. Lattice points denote atomic sites and black (red) regions the scattering region (leads). The device width is 25 nm while the width of leads 2, 3, 5 and 6 and the interlead separation is 20 nm. The tight-binding parameters in units of $t$ are: $t_\perp = 0.13$, $\lambda_{_{VZ}} = 0.08$, $\lambda_R = 0.06$, $M = 0.08$, $U = 0.004$.} 
\label{fig_F4}
\end{figure}

In virtue of the bulk-boundary correspondence, a bulk topological gap implies the existence of edge states in a finite size system. It is thus instructive to scrutinize the band structure of a finite BG sample, as plotted in Fig. \ref{fig_F4}(a) for a zigzag ribbon of width 25 nm. There are several in-gap states that correspond to edge states \cite{Frank2018, Island2019}. The two pair of crossing bands in the middle of the Brillouin zone are trivial edge states arising from the valley-Zeeman SOC \cite{Gmitra2016, Frank2018}, while the bands in the right valley are also non-topological edge states that appear in a zigzag BG with nonzero $U$ \cite{Qiao2011}. Differently, the bands inside the gap of the left valley are the ones expected in the QAHE phase, as highlighted by the red color in Fig. \ref{fig_F4}(a). Therefore, there are at least three propagating modes in the bulk gap, although in principle only one should convey the Hall response. To corroborate this, we carry out quantum transport simulations in a Hall-bar device (inset in Fig. \ref{fig_F4}(b)) to evaluate the transverse Hall voltage as a response to a longitudinal current \cite{methods}. 

The Hall resistance versus energy is plotted in Fig. \ref{fig_F4}(b) and shows a clear quantization at $h/e^2$ in the energy region inside the bulk band gap seen in Fig. \ref{fig_F4}(a) (shaded gray region). This indicates the presence of a VP-QAHE and confirms that only one state contributes to edge conduction. We note that at energies below $E=0$ the Hall resistance is not quantized (but still finite) because one of the bulk bands overlaps with the edge state, which likely occurs due to the large value of $\lambda_R$ that slightly distorts the bulk band dispersion \cite{Rashba}.

Next, we add static disorder to the Hamiltonian $\mathcal{H}$ to test the robustness of these states. We introduce (i) bulk Anderson disorder with the term $H_A = \sum_i W_i c^\dagger_{is} c_{is}$, where $W_i \in [-W/2, W/2]$ is a random potential uniformly distributed at each site $i$, and (ii) edge disorder, which is modeled with two parameters \cite{Mucciolo2009, Caridad2018}: the probability $P$ and the number of sweeps, $S$ (the resulting disordered edges are depicted in the inset of Fig. \ref{fig_F4}(b)). For each sweep $i$, a random number $p \in [0,1]$ is assigned to each atom at the edge of the device and such site is removed if $p < P$. The several plots in Fig. \ref{fig_F4}(b) show that both types of disorder do not significantly affect $R_H$. For bulk disorder, even when the disorder strength is more than 30 times larger than the band gap, the quantization is still resilient in a given energy window. For edge disorder, the effect is minimal even though we use $P = 0.8$ and $S=10$, which corresponds to having removed on average $\sum_i^S P \sim 8$ unit cells on the edges.  

To discuss the possible measurement of VP-QAHE,  we plot in Fig. \ref{fig_F5} the Hall conductivity for parameters extracted from experiments or {\it ab initio} simulations, namely, $\lambda_{_{VZ}} = \lambda_R  = 2$ meV for graphene/transition metal dichalcogenide heterostructures with proper alignment or twist angle \cite{Li2019, David2019, Naimer2021} and $M \approx 30$ meV for graphene on CrSe \cite{Wu2020} or on CrSBr \cite{Ghiasi2021}. The plateau extends over 0.5 meV and is visible below $T = 1$ K. For a larger value of the Rashba SOC of 10 meV \cite{Island2019}, the effect persists up to 2 K with a plateau width of $\approx 2$ meV (see inset in Fig. \ref{fig_F5}).

In this work, the formation of a fully valley-polarized QAHE has been predicted in BG when magnetic and SOC proximity effects are imprinted on different layers. Such topological phase was previously suggested in silicene \cite{Pan2014, Pan2015} and in Cu-decorated In lattices on Si(111) surfaces \cite{Zhou2017}. Bilayer graphene is however more chemically stable and easier to integrate in electronic devices, and it further allows tailoring proximity effects by properly choosing the substrates or twist angles between layers. Additionally, a correlated-induced VP-QAHE has also been observed in several moir\'e superlattices such as twisted bilayer graphene \cite{Serlin2020}, trilayer graphene/hexagonal boron nitride \cite{Chen2020}, twisted monolayer/bilayer graphene \cite{Polshyn2020} or twisted bilayer graphene aligned with hexagonal boron nitride \cite{Tschirhart2021}. In these systems, the electron correlation appearing at flat bands leads to a valley-polarization that breaks time-reversal symmetry, and together with a large Berry curvature of the valley states, produces a VP-QAHE at some band fillings. This mechanism thus requires strong correlations induced by twisting adjacent layers, while our mechanism relies on the proximity effect of adjacent layers. It is worth mentioning that the recent experimental advances in twist angle engineering and control could be used in the context of our proposal to fine-tune both the spin-orbit \cite{Li2019, David2019, Naimer2021} and exchange proximity effects \cite{Zollner2021}. 

In conclusion, our simulations reveal that the edge states are strongly robust to both bulk and edge disorders, therefore offering a way to create persistent valley-polarized currents, in contrast to the vast majority of valley-related phenomena where valley currents are fragile and very sensitive to short-range scatterers \cite{Cresti2016}. Both the valley-polarization and chirality of the VP-QAHE can be reversed by changing the magnetization of the magnetic material, and a perpendicular electric field drives the topological phase transition. This could be used to create topological field-effect transistors \cite{Qian2014} or valley filters by combining different regions with opposite magnetizations. Overall, the potential to induce a layer-dependent interaction such as spin-orbit, (anti)ferromagnetic or superconducting \cite{Rio2021} proximity effect puts forward bilayer graphene as a promising material to create versatile topological phases for next-generation electronics.

\begin{acknowledgments}
We thank Aron W. Cummings for helpful comments. The authors were supported by the European Union Horizon 2020 research and innovation programme under Grant Agreement No. 881603 (Graphene Flagship) and  No. 824140  (TOCHA, H2020-FETPROACT-01-2018). ICN2 is funded by the CERCA Programme/Generalitat de Catalunya, and is supported by the Severo Ochoa program from Spanish MINECO (Grant No. SEV-2017-0706).
\end{acknowledgments}


\begin{thebibliography}{102}
\bibitem [{H0()}]{H0}%
  \BibitemOpen
  \href@noop {} {}\bibinfo {note} {The orbital part of the Hamiltonian is
  $\mathcal{H}_0 = -t \sum_{\langle i,j \rangle, s} c^\dagger_{is} c_{js} +
  t_\perp \sum_{\langle i\in A_1, j\in B_2 \rangle, s} c^\dagger_{is} c_{js} +
  U \sum_{i,s} \eta_i c^\dagger_{is}c_{is}$. Here, $c_{is}^\dagger$ ($c_{is}$)
  is the creation (annihilation) operator of electrons at site $i$ with spin
  $s$, $t$ is the in-plane nearest-neighbor hopping within a layer and
  $t_\perp$ the interlayer hopping that connects vertically-stacked sites
  belonging to the A sublattice in layer 1 with B atoms in layer 2. The last
  term in $\mathcal{H}_0$ describes the layer potential $U$ that appears due to
  an interlayer asymmetry such as an external perpendicular electric field. The
  sign of $U$ is opposite for different layers, as indicated by $\eta_i = \pm
  1$ for sites belonging to layer 1 and 2, respectively. Additional interlayer
  terms are found to bring no qualitative new features in the results
  \cite{Suppmat}.}\BibitemShut {Stop}%
\bibitem [{Ras()}]{Rashba}%
  \BibitemOpen
  \href@noop {} {}\bibinfo {note} {For computational convenience, here large
  enough values of $\lambda_R$ are used to generate a topological gap and
  observing edge states in small width nanoribbons.}\BibitemShut {Stop}%
\end{thebibliography}

%

\end{document}